# Extinction of guided light induced by coupled spiral meta-atom resonators at arbitrary order exceptional points

Chengzhi Zhang[1,2] and Shubo Wang[1,3,*]

[1]Department of Physics, City University of Hong Kong, Tat Chee Avenue, Kowloon, Hong Kong, China
[2]Tsinghua Shenzhen International Graduate School, Tsinghua University, Shenzhen, Guangdong 518055, China.
[3]City University of Hong Kong Shenzhen Research Institute, Shenzhen, Guangdong 518057, China
*Corresponding author: shubwang@cityu.edu.hk

## Abstract

Exceptional points (EPs) in non-Hermitian systems can give rise to intriguing effects not available in conventional Hermitian systems due to their unusual properties. Using full-wave simulations, we investigate the scattering, absorption, and transmission of guided light at arbitrary order exceptional points in a non-Hermitian system consisting of coupled spiral meta-atom resonators sitting on a dielectric waveguide. The EPs are realized by exploiting the unidirectional coupling of the chiral dipole modes in the subwavelength meta-atom resonators. The scattering and absorption of the resonators induce the attenuation of the guided light in the dielectric waveguide. We show that the EPs can give rise to a plateau in the attenuation spectrum of the guided light with near-zero transmission, i.e., the guided light is almost completely dissipated via the resonators in the forms of intrinsic material loss and radiation loss. In addition, the width of this plateau (i.e., the extinction bandwidth) increases with the order of the EP. The phenomena can be understood by employing a coupled mode analysis, with the analytical results quantitatively agree with the numerical results. The results may find applications in designing novel on-chip optical absorbers and sensors.

## 1. Introduction

Exceptional points (EPs) of non-Hermitian Hamiltonians, at which eigenvalues and their relevant eigenvectors coalesce simultaneously [1–5], can give rise to many interesting phenomena and applications, such as enhanced optical isolation [6,7], unidirectional light propagation [8], optical switches [9-11], circular polarizer [12], unconventional lasers [5,13,14], non-Hermitian skin effects [15], asymmetric mode conversion [16], and topological phases [17], etc. Recently, the absorption and transmission properties of guided light at EPs have drawn considerable attention, the study of which has uncovered intriguing phenomena such as chiral perfect absorption [18,19]. The chiral perfect absorption can give rise to an unusual resonance line shape of the absorption/transmission spectrum different from that of conventional systems. The phenomenon is attributed to the critical coupling of the resonators and the chirality of the EPs. The previous studies are restricted to the lowest order EP due to the challenge of realizing higher order EPs, which usually requires tuning multiple system parameters in a large parameter space. The absorption and transmission properties of guided light at higher order EPs remains largely unexplored.

A simple and robust mechanism to realize higher order EPs is the unidirectional coupling between multiple resonators. For two coupled resonators described by a non-Hermitian Hamiltonian $H = \begin{bmatrix} a & c \\ d & b \end{bmatrix}$, if $a = b$ and either $c = 0$ or $d = 0$, the eigenvalues and eigenvectors naturally coalesce, corresponding to a second order EP [20,21]. There are several approaches that can be applied to realize unidirectional coupling, including the interferences of different scattering channels [20,22] and the spin-momentum locking of guided waves [23–27]. The latter is particularly interesting since it originates from an intrinsic property of the evanescent field in a guided wave, i.e., the direction of its spin is locked to the propagating direction (i.e., direction of linear momentum) of the associated guided wave. The robustness of this approach has been utilized to realize arbitrary order exceptional surface in coupled particles [21,28], which gives rise to a light funneling phenomenon similar to the non-Hermitian skin effect [29]. Here, we implement this approach in a different structure to investigate the scattering, absorption and transmission properties of guided light at arbitrary order EPs.

In this paper, we consider a two-dimensional non-Hermitian system consisting of multiple spiral meta-atom resonators sitting above a dielectric waveguide. We show that the chiral dipole mode, comprising both circularly polarized electric dipole and toroidal dipole, of the meta-atoms can unidirectionally couple with each other via the dielectric waveguide due to the spin-momentum locking of the guided wave. This gives rise to an EP with its order decided by the number of the meta-atoms. We show that the strong coupling between the meta-atoms and the waveguide induces near-zero transmission of the guided light at the resonance frequency of the chiral dipole mode, and the input light energy is almost completely dissipated by the meta-atoms via material loss and radiation loss, corresponding to near-perfect extinction of the input light. Interestingly, the order of the EP directly decides the bandwidth of near-perfect extinction, i.e., the bandwidth is larger for a higher order EP. We apply a coupled mode analysis to uncover the physics behind the phenomena and obtain analytical results consistent with the full-wave numerical results.

The paper is organized as follows. In Section 2, we present the numerical results of the scattering and absorption of guided light by a single spiral meta-atom. In Section 3, we present the numerical and analytical results for the light extinction induced by coupled spiral meta-atoms at the EPs of different orders, and we discuss the underlying physics. Finally, we draw the conclusion in Section 4.

## 2. Extinction of guided light induced by a spiral meta-atom

We first consider a spiral resonator in free space under the incidence of a plane wave with electric field linearly polarized in $x$ direction, as shown in Fig. 1(a). The spiral resonator has a Swiss-roll structure with four layers and a radius of $r = 8$ mm. We assume the boundary (denoted by the black lines) is perfect electric conductor (PEC), and the resonator is filled with lossy dielectric material with relative permittivity $\varepsilon = 2 + i\gamma$. The structure is excited by a plane wave with wavevector along $-y$ direction and electric field polarized linearly along $x$ direction. We conduct full-wave simulations of the structure by using COMSOL Multiphysics and calculate its absorption cross section as a function of the material loss $\gamma$ and excitation frequency, assuming an open boundary condition in the far field. As shown in Fig. 1(b), the absorption cross section changes with the frequency and material loss, and its value can reach the single resonance limit $\lambda/2\pi$ at the resonance frequency $f = 0.673$ GHz. Figure 1(c) shows the electric and magnetic dipole moments induced in the resonator. As seen, the magnetic dipole moment dominates in the response of the resonator. The magnetic field $|H_z|$ at the resonance frequency is shown in Fig. 1(d), which shows a pattern of magnetic monopole with strong magnetic field localized inside the resonator. This magnetic resonance is of Fabry-Perot type arising from the propagation and reflection of TM guided wave inside the layers of the spiral resonator [30].

To construct a dipole particle for use in realizing arbitrary order EPs, we consider the meta-atom composed of four spiral resonators arranged symmetrically on a circle with diameter $D = 78$ nm, as shown in Fig. 2(a). We put the meta-atom above a dielectric slab waveguide made of silicon with relative permittivity 12. The distance between the center of the meta-atom and the upper surface of the waveguide is $h$. The TM guided wave is excited at the left end of the waveguide and propagates to the right side. We calculate the reflection and attenuation of the guided wave as a function of the frequency and $h$ with $\gamma = 2 \times 10^{-5}$, as displayed in Fig. 2(b) and Fig. 2(c), respectively. The attenuation is defined as $A = 1 - T - R$, where $T$ is the power transmission and $R$ is the power reflection. We note that the power attenuation includes both the power absorbed (corresponding to material loss) and the power scattered (corresponding to radiation loss) by the meta-atom. As seen, the reflection in Fig. 2(b) nearly vanishes in the considered frequency range (near the resonance frequency), and the attenuation in Fig. 2(c) is strongly enhanced with the maximum value approaching 100%. We calculate the multipole moments induced in the meta-atom. Specifically, the electric dipole $\mathbf{p}$, the magnetic dipole $\mathbf{m}$, and the toroidal dipole $\mathbf{t}$ can be expressed as [31-33]:

$$\mathbf{p} = \frac{1}{i\omega} \int \mathbf{j} d^3 r,$$
$$\mathbf{m} = \frac{1}{2} \int (\mathbf{r} \times \mathbf{j}) d^3 r, \tag{1}$$
$$\mathbf{t} = \frac{1}{10} \int [(\mathbf{r} \cdot \mathbf{j})\mathbf{r} - 2r^2 \mathbf{j}] d^3 r,$$

where $\mathbf{r}$ is the position vector, $\mathbf{j}$ is the current density. Figure 2(d) shows the obtained dipole moments. As seen, the electric dipole and toroidal dipole dominate near the resonance frequency. Figure 2(e) shows the relative amplitude and phase of the electric dipole components $p_x$ and $p_y$, and the relative amplitude and phase of the toroidal dipole components $t_x$ and $t_y$. At the resonance frequency, $p_x$ and $p_y$ have approximately equal amplitude and a relative phase difference of $\pi/2$. This is also true for $t_x$ and $t_y$. Consequently, the induced electric dipole and toroidal dipole are both nearly circular polarized, and the resonance mode of the meta-atom can be considered a chiral dipole mode. The fields of these circular dipoles only couple to the guided wave propagating in $+x$ direction due to the spin-momentum locking of the guided wave.

The rate equation for the chiral dipole mode of the meta-atom can be written as [21, 34, 35]:

$$\frac{da}{dt} = -i\omega_0 a - \frac{\gamma_1 + \gamma_c}{2} a - \sqrt{\gamma_c} a_{in}, \tag{2}$$

where $\omega_0$ denotes the resonance frequency of the meta-atom; $\gamma_1$ and $\gamma_c$ represent the material loss and radiation loss of the meta-atom, respectively; $a$ denotes the mode amplitude; $a_{in}$ represents the amplitude of the incident guided wave. We can solve the rate equation for the steady state, and the mode amplitude can be expressed as:

$$a = \frac{\sqrt{\gamma_c}}{i\Delta - \Gamma} a_{in}, \tag{3}$$

where $\Gamma = (\gamma_1 + \gamma_c)/2$ and $\Delta = \omega - \omega_0$ denotes the frequency detuning of incident wave. The chiral dipole mode can be coupled to the guided wave propagating along $+x$ direction. The amplitude of the output guided wave can be written as:

$$a_{out} = (a_{in} + \sqrt{\gamma_c} a + b a_{in}) = \left(1 + b + \frac{\gamma_c}{i\Delta - \Gamma}\right) a_{in}, \tag{4}$$

where $b$ is a complex number denoting the contribution of the other modes spectrally away from the resonance frequency. And, $b = 0$ corresponds to the ideal case where the output guided wave is contributed only by the chiral dipole mode of the meta-atom in the considered frequency range. Using Eq. (4), the normalized transmission $T$ and attenuation $A$ for the intensity of the incident guided light can be expressed as:

$$T = \left|\frac{a_{out}}{a_{in}}\right|^2 = \left|1 + b + \frac{\gamma_c}{i\Delta - \Gamma}\right|^2,$$
$$A = 1 - T = 1 - \left|1 + b + \frac{\gamma_c}{i\Delta - \Gamma}\right|^2. \tag{5}$$

Here, we have neglected the reflection which is close to zero near the resonance frequency. According to Eq. (5), when $b = 0$ and $\gamma_1 = \gamma_c$, corresponding to critical coupling, we have $T = 0$ and $A = 1$ at the resonance frequency $\omega = \omega_0$ [19]. We calculate the transmission and attenuation for the system with $\gamma = 2 \times 10^{-5}$ and $h = 30.3$ mm, as depicted by Fig. 2(f). The dot lines represent the results of full-wave simulations. The solid lines are obtained via fitting the numerical results with Eq. (5). We find excellent agreement between them. The attenuation spectrum shows a Lorentzian shape due to the resonance of the meta-atom, which approaches 100% at the resonance frequency. Meanwhile, the transmission shows a dip close to zero. We further calculate the power dissipated in the meta-atom and the power scattered by the meta-atom, denoted by the green dot line and black dot line, respectively. As seen, the scattering loss (i.e., radiation loss of the meta-atom) dominates at the resonance frequency.

## 3. Extinction of guided light at exceptional points of different orders

We consider two identical meta-atoms on the waveguide with distance $d$ = 3700 mm, shown in Fig. 3(a). Using the coupled mode theory (CMT), the rate equations for the two coupled meta-atoms can be written as [21, 34, 35]:

$$\frac{da_1}{dt} = -i\omega_0 a_1 - \frac{\gamma_1 + \gamma_c}{2} a_1 - i\kappa_{12} a_2 - \sqrt{\gamma_c} a_{in1}, \qquad (6)$$
$$\frac{da_2}{dt} = -i\omega_0 a_2 - \frac{\gamma_2 + \gamma_c}{2} a_2 - i\kappa_{21} a_1 - \sqrt{\gamma_c} a_{in2},$$

where $\gamma_1 = \gamma_2$; $\kappa_{21}$ ($\kappa_{12}$) denotes the coupling parameter from meta-atom 1 (2) to meta-atom 2 (1); $a_1$ and $a_2$ denote the mode amplitudes of meta-atom 1 and meta-atom 2, respectively; $a_{in1}$ and $a_{in2}$ represents the incident wave amplitudes at the position of the meta-atoms with the relation $a_{in2} = a_{in1} \exp(ik_{wg}d)$, where the $k_{wg}$ denotes the propagation constant of the guided wave. The above equations can be re-written as:

$$\frac{d\Lambda}{dt} = -iH\Lambda - \sqrt{\gamma_c}\Lambda_{in}, \qquad (7)$$

and

$$H = \begin{bmatrix} \omega_0 - i\Gamma & \kappa_{12} \\ \kappa_{21} & \omega_0 - i\Gamma \end{bmatrix}, \Lambda = \begin{bmatrix} a_1 \\ a_2 \end{bmatrix}, \Lambda_{in} = \begin{bmatrix} a_{in1} \\ a_{in2} \end{bmatrix}. \qquad (8)$$

Since the chiral dipole mode of the meta-atoms only couples to the guided wave propagating in +x direction, we have $\kappa_{12} = 0$ and $\kappa_{21} \neq 0$, corresponding to unidirectional coupling between the two meta-atoms. This naturally gives rise to a second order EP of the Hamiltonian in Eq. (8). Assuming the time-harmonic mode incident field $a_{in1} = a_{in}$, the mode amplitude of the meta-atom 1 can be expressed as $a_1 = a_{in}\sqrt{\gamma_c}/(i\Delta - \Gamma)$, and the guided wave at the position of meta-atom 1 can be obtained as [19]:

$$a_{out1} = (a_{in} + \sqrt{\gamma_c}a + ba_{in}) = \left(1 + b + \frac{\gamma_c}{i\Delta - \Gamma}\right) a_{in}, \qquad (9)$$

which has the same expression as in the case of Fig. 2(a). Since the two particles are identical, we can straightforwardly derive the amplitude $a_{out2}$ at the position of meta-atom 2 as

$$a_{\text{out2}} = \left(1 + b + \frac{\gamma_c}{i\Delta - \Gamma}\right) a_{\text{out1}} = \left(1 + b + \frac{\gamma_c}{i\Delta - \Gamma}\right)^2 a_{\text{in}}. \tag{10}$$

Therefore, the transmission and attenuation of the guided wave at the right end of the waveguide can be written as:

$$T = \left|\frac{a_{\text{out2}}}{a_{\text{in}}}\right|^2 = \left|1 + b + \frac{\gamma_c}{i\Delta - \Gamma}\right|^4,$$
$$A = 1 - T = 1 - \left|1 + b + \frac{\gamma_c}{i\Delta - \Gamma}\right|^4. \tag{11}$$

We conduct full wave simulations and calculate the transmission $T$ and attenuation $A$ for this case. The results are shown in Fig. 3(b) by the dot lines. The fitting results with Eq. (11) are denoted by the solid lines, which agree well with the numerical results. Notably, the attenuation and transmission have a quartic lineshape near the resonance frequency, as predicted by our analytical model based on the CMT. Thus, the emergence of the EP leads to an enlarged width of the resonance peak (dip) in the attenuation (transmission).

Higher-order EPs can be directly realized in our system by putting more identical spiral meta-atoms on the slab waveguide. For $N$ meta-atoms unidirectionally coupled through the waveguide, the rate equations can be expressed as [21,34,35]:

$$\frac{da_i}{dt} = -i\omega_0 a_i - \Gamma a_i - i\kappa_{ij} \sum_{j=1}^{i-1} a_j - \sqrt{\gamma_c} a_{\text{in}i}. \tag{12}$$

The corresponding effective Hamiltonian is

$$H = \begin{bmatrix} \omega_0 - i\Gamma & 0 & \cdots & 0 \\ \kappa_{21} & \omega_0 - i\Gamma & \cdots & 0 \\ \vdots & \vdots & \ddots & \vdots \\ \kappa_{N1} & \kappa_{N2} & \cdots & \omega_0 - i\Gamma \end{bmatrix}, \tag{13}$$

where the non-zero coupling parameters $\kappa_{ij}$ with $i > j$ are attributed to the unidirectional coupling among the meta-atoms. The $N$ complex eigenvalues are degenerated at $\omega = \omega_0 - i\Gamma$ which corresponds to a $N^{\text{th}}$ order EP. The transmission and attenuation of the guided wave at the EP can be written as:

$$T = \left|\frac{a_{\text{out}N}}{a_{\text{in}}}\right|^2 = \left|1 + b + \frac{\gamma_c}{i\Delta - \Gamma}\right|^2 \left|\frac{a_{\text{out}N-1}}{a_{\text{in}}}\right|^2 = \left|1 + b + \frac{\gamma_c}{i\Delta - \Gamma}\right|^{2N},$$
$$A = 1 - T = 1 - \left|1 + b + \frac{\gamma_c}{i\Delta - \Gamma}\right|^{2N}, \tag{14}$$

where $a_{\text{out}N}$ and $a_{\text{out}N-1}$ denote the guided wave amplitudes at the particle *N* and *N*-1, respectively. We notice that the attenuation *A* and transmission *T* are functions with a power of $2N$. Therefore, an increase of the EP order leads to an enlarged bandwidth of the near-perfect attenuation and near-zero transmission in the spectra.

To illustrate the high-order EP effect, we conduct full wave simulations for the systems with *N* = 3 and *N* = 4 meta-atoms, which give rise to the third and fourth order EPs, respectively. The system with *N* = 4 is shown in Fig. 4(a). The numerical results for the attenuation and transmission are represented by the dot lines in Fig. 4(b). The analytical results of CMT obtained by fitting the numerical results with Eq. (14) are denoted by the solid lines. We notice a quantitative agreement between the two. Importantly, the width of the plateau in the attenuation spectrum gradually increases when the order of the EP increases, as expected from the analysis based on the CMT. This demonstrates the validity of our analytical model and interpretation based on the unidirectional coupling of the meta-atoms.

## 4. Conclusion

To summarize, we realize arbitrary order EPs in spiral meta-atom resonators strongly coupled via a dielectric waveguide. This is achieved by employing the unidirectional coupling among the meta-atoms protected by the spin-momentum locking of the guided wave. We demonstrate that the EPs can give rise to enhanced attenuation of the guided light. In contrast to previous studies employing whispering gallery modes to achieve perfect absorption at the second order EP, the proposed system employs chiral dipole modes comprising circular electric dipole and toroidal dipole to realize near-perfect extinction of guided light at higher order EPs. Using full-wave numerical simulations and CMT, we show that a higher order EP can give rise to a larger extinction bandwidth. Our study contributes to the understanding of light-matter interactions in non-Hermitian strongly coupled systems. The results may find applications in on-chip light absorption and optical sensing.

## Acknowledgements

This work was supported by the Research Grants Council of the Hong Kong Special Administrative Region, China (Project Nos. CityU 11301820 and AoE/P-502/20) and National Natural Science Foundation of China (11904306, 12322416).

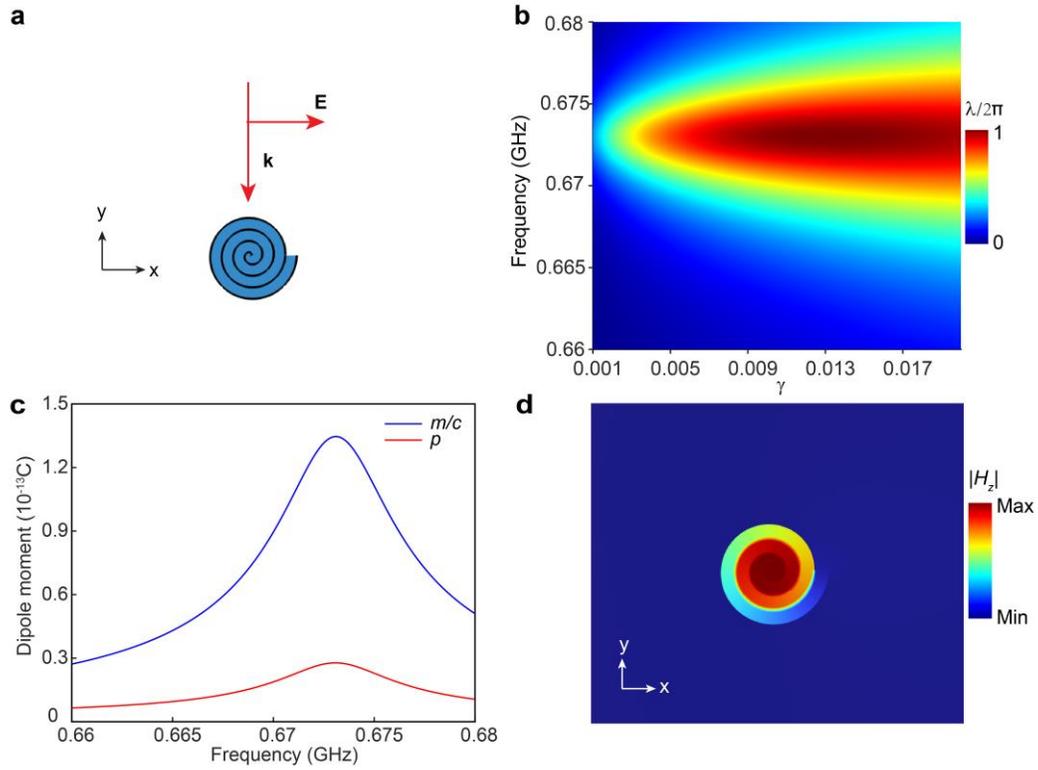

Fig. 1. (a) The spiral resonator under plane wave excitation. (b) Absorption cross section of the spiral resonator as a function of frequency and material loss $\gamma$. (c) Electric and magnetic dipole moments of the resonator as a function of frequency. (d) $|H_z|$ field at the resonance frequency $f = 0.673$ GHz.

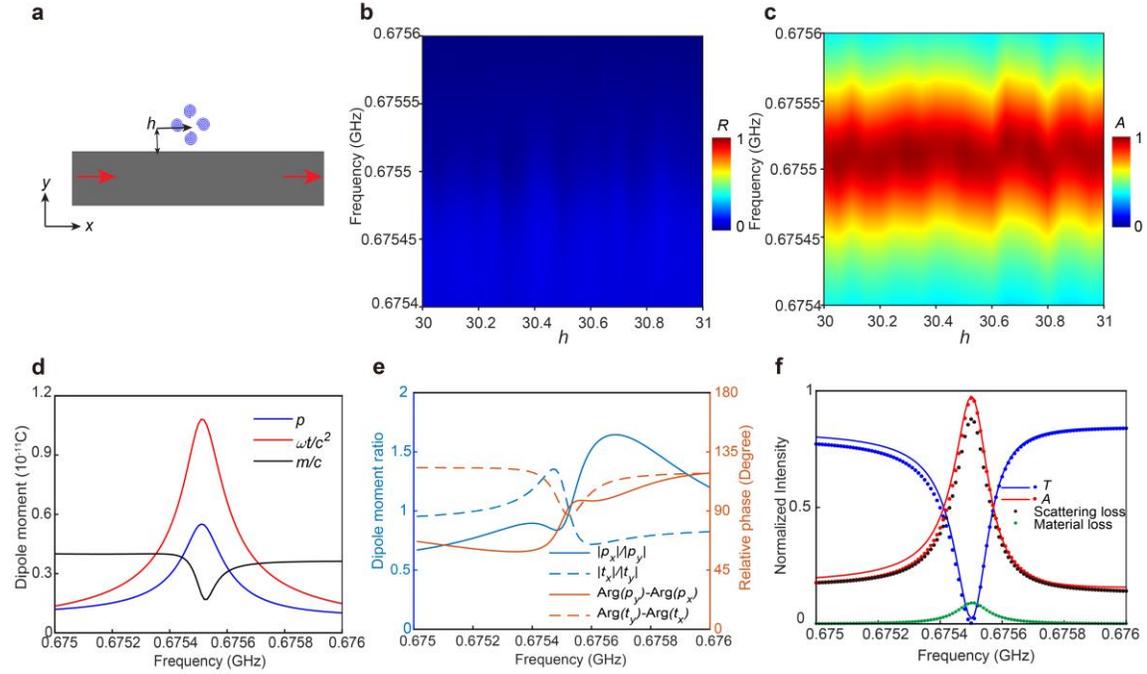

Fig. 2. (a) Spiral meta-atom sitting above a slab waveguide under port excitation. (b) Reflection as a function of the height $h$ and frequency. (c) Attenuation as a function of the height $h$ and frequency. (d) Dipole components of the meta-atom as a function of frequency. (e) Relative amplitude and phase of the electric dipole components $p_x$ and $p_y$, and relative amplitude and phase of the toroidal dipole components $t_x$ and $t_y$. We set $\gamma = 2 \times 10^{-5}$ and $h$ = 30.3 mm. (f) Numerical results (blue/red dot lines) and analytical fitting results (bule/red solid lines) of transmission and attenuation as a function of frequency. The black dot line denotes the scattering loss of the meta-atom. The green dot line denotes the intrinsic material loss of the meta-atom.

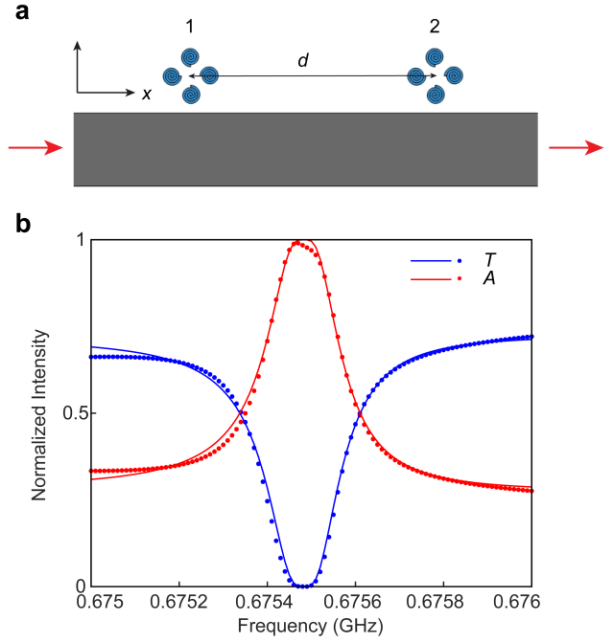

Fig. 3. (a) Two meta-atoms sitting above the dielectric slab waveguide with distance $d = 3700$ mm and under port excitation. (b) Numerical results (dot lines) and CMT fitting results (solid lines) of transmission and attenuation.

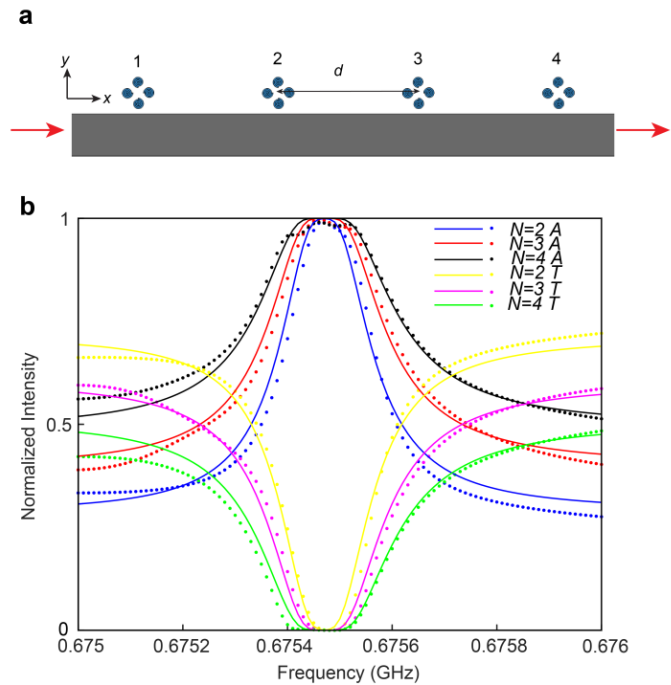

Fig. 4 (a) Four meta-atoms sitting above the dielectric slab waveguide with distance $d = 3700$ mm and under port excitation. (b) Numerical results (dot lines) and CMT fitting results (solid lines) of transmission and attenuation as a function of frequency.